\documentclass[runningheads]{llncs}
\usepackage[T1]{fontenc}
\usepackage{lipsum}
\usepackage{rotating}
\usepackage{makecell} 
\usepackage{tabularx,booktabs,ragged2e}
\usepackage{array}
\usepackage{float}
\newcolumntype{L}{>{\RaggedRight\arraybackslash}X}

\newcolumntype{C}[1]{>{\centering\let\newline\\\arraybackslash\hspace{0pt}}m{#1}}
\usepackage[section]{placeins}
\usepackage{graphicx}

\usepackage[hidelinks]{hyperref}

\begin{document}
    \title{A Hybrid Encryption Framework Combining Classical, Post-Quantum, and QKD Methods}
    \titlerunning{A Hybrid Encryption with Classical, PQC, and QKD}
    %
    \author{Amal Raj\inst{1}\orcidID{0009-0001-4025-1045} \and
    {Vivek Balachandran\inst{2}\orcidID{0000-0003-4847-7150}
    }}
    %
    \authorrunning{A. Raj and V. Balachandran}
    %
    \institute{Research Engineer, Singapore Institute of Technology \\ \email{amal.raj@singaporetech.edu.sg} \and
    Associate Professor, Singapore Institute of Technology  \\
    \email{vivek.b@singaporetech.edu.sg}\thanks{Corresponding author}}
    \maketitle 

    \begin{abstract}
        This paper introduces a hybrid encryption framework combining classical cryptography (EdDSA, ECDH), post-quantum cryptography (ML-DSA-6x5, ML-KEM-768), and Quantum Key Distribution (QKD) via Guardian to counter quantum computing threats. Our prototype implements this integration, using a key derivation function to generate secure symmetric and HMAC keys, and evaluates its performance across execution time and network metrics. The approach improves data protection by merging classical efficiency with PQC’s quantum resilience and QKD’s key security, offering a practical transition path for cryptographic systems. This research lays the foundation for future adoption of PQC in securing digital communication.
        
        \keywords{Hybrid Encryption \and Post-Quantum Cryptography \and Quantum Key Distribution}
    \end{abstract}

    \section{Introduction}
        Digital communication has long been protected by classical encryption, which uses algorithms such as Elliptic Curve Cryptography (ECC) \cite{ecc} and Rivest-Shamir-Adleman (RSA) \cite{rsa} to guarantee security in a variety of applications. These systems are based on intricate mathematical problems like discrete logarithms and integer factorization. However, quantum computing threatens traditional cryptosystems, as algorithms such as Shor’s \cite{shor} can efficiently solve their underlying mathematical problems. In a time when quantum computers are becoming more and more practical, this impending threat calls for a move toward quantum-resistant solutions to safeguard sensitive data.

        The first quantum revolution in the $20^{th}$ century laid the foundation for quantum physics, revealing phenomena such as wave-particle duality, while the second revolution, currently underway, advances fields such as quantum information and optics \cite{ref-1}. Theoretically immune to eavesdropping, Quantum Key Distribution (QKD) provides secure key exchange based on quantum mechanics; however, practical constraints like cost, distance, and key rates prevent its widespread use \cite{ref-3,ref-4}. Post-quantum cryptography (PQC) is a promising field that aims to design algorithms resistant to quantum attacks. Since fully replacing classical methods is impractical, we propose a hybrid encryption model combining classical methods like Elliptic Curve Diffie-Hellman (ECDH) \cite{ecdh} with PQC techniques such as Module-Lattice-based Key Encapsulation Mechanism (ML-KEM) \cite{ml-kem} and QKD via Guardian \cite{guardian}.
        
    \section{Proposed Method}
\begin{figure}[htb!]
                \centering
                \includegraphics[width=1\linewidth]{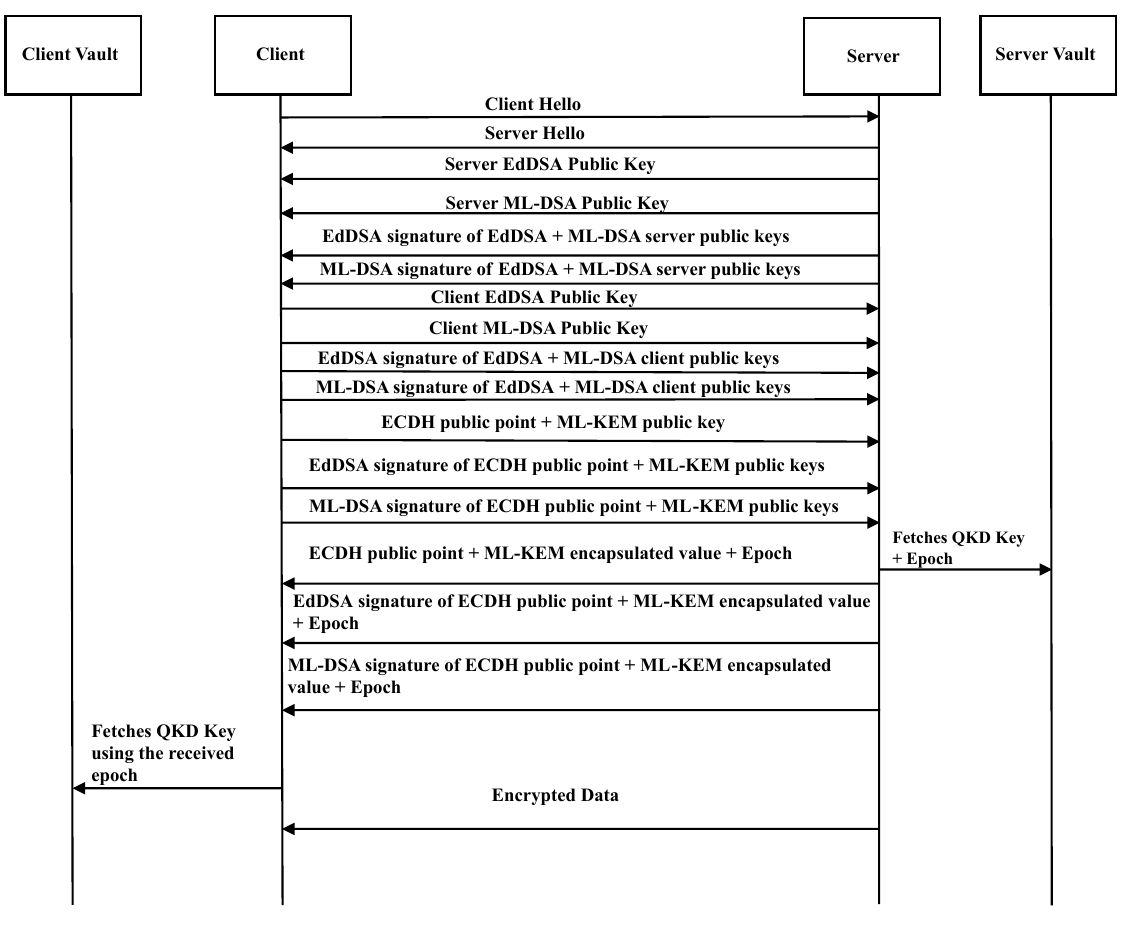}
                \caption{Hybrid Encryption Framework Flowchart}
                \label{fig:encryption-flowchart}
            \end{figure}

        This project develops a hybrid encryption framework to secure communication between two nodes, integrating classical cryptographic techniques with post-quantum cryptography (PQC) and Quantum Key Distribution (QKD), as shown in Fig. \ref{fig:encryption-flowchart}. The protocol employs a hybrid key exchange scheme that combines the classical ECDH protocol using the X25519 curve, the post-quantum ML-KEM with ML-KEM-768, and QKD managed by Guardian. These keys are concatenated and passed through a key derivation function (KDF) to produce symmetric encryption and HMAC keys, ensuring security even if one protocol is compromised, as their generation depends on all three components (Fig. \ref{fig:encryption-flowchart}). For signatures, a double signature scheme uses the classical Edwards-curve Digital Signature Algorithm (EdDSA) \cite{edDSA} with Ed25519 and the post-quantum Module-Lattice-based Digital Signature Algorithm (ML-DSA) \cite{ml-dsa} with ML-DSA-6x5 in parallel, requiring both signatures for validation to ensure resilience against classical and quantum attacks. Guardian retrieves quantum-generated keys from a simulator or Vault, synchronized via epochs, and distributes them across multi-node networks using ETSI TS QKD 014-compliant APIs, enhancing scalability. This framework delivers a practical prototype, evaluated through performance metrics, to protect data against quantum threats.
        
        \subsection{Implementation Details}
            The hybrid framework leverages multiple key exchange implementations, combining classical X25519 ECDH with post-quantum KEMs for robust security. These include FrodoKEM (standard and enhanced variants of FrodoKEM, with SHAKE and AES modes), ML-KEM (using X25519/X448 curves and security levels 512, 768, 1024), Kyber-768, and McEliece \cite{mceliece}. Each implementation generates shared secrets through ECDH and the respective KEM, employing KDF2 (SHA-256) for key derivation and, except for FrodoKEM, a random number generator to ensure resilience against classical and quantum attacks. Implementation of this framework is available in the GitHub repository mentioned in \cite{pqchybrid}.

            \FloatBarrier
            
        \subsection{Performance Evaluation Framework} \label{performance-eval}
            A performance evaluation framework was created to measure the time taken to implement each of the PQC schemes evaluated in Table \ref{tab:performance-framework}, as well as the number of packets needed for transmission. The measured metrics include \textit{average execution time}, \textit{standard deviation}, \textit{maximum and minimum execution time}, together with the \textit{number of bytes transferred} and \textit{packets transferred}. Cryptographic strength estimation is a critical aspect in comparing different key exchange methods. For classical cryptographic methods such as ECDH (X25519), strength is typically determined by the key length and the difficulty in solving the discrete logarithm problem. For PQC methods, such as those in Table \ref{tab:performance-framework}, strength is evaluated by the computational complexity of solving hard mathematical problems resistant to quantum computing attacks.
            
            \begingroup
                \renewcommand{\arraystretch}{1.1}
                \setlength{\tabcolsep}{0pt}
                \renewcommand\tabularxcolumn[1]{>{\centering\arraybackslash}m{#1}}  
                \newcolumntype{P}[1]{>{\centering\arraybackslash}m{#1}}
                
                \begin{table}[h!]
                    \centering
                    \caption{Performance Metrics for PQC Implementations Across Evaluated Schemes}
                    {\scriptsize
                        \begin{tabularx}{\textwidth}{|@{\hspace{1pt}}P{0.14\textwidth}@{\hspace{7pt}}|@{\hspace{1pt}}P{0.11\textwidth}@{\hspace{5pt}}|P{0.11\textwidth}@{\hspace{5pt}}|@{\hspace{1pt}}P{0.09\textwidth}@{\hspace{6pt}}|@{\hspace{1pt}}P{0.08\textwidth}@{\hspace{5pt}}|@{\hspace{1pt}}P{0.05\textwidth}@{\hspace{6pt}}|@{\hspace{-2pt}}X@{\hspace{-3pt}}|X|X|}
                        \hline
                            \makecell{Function} & 
                            \makecell{Average\\(ns)} & 
                            \makecell{Std Dev\\(ns)} & 
                            \makecell{Max\\(ns)} & 
                            \makecell{Min\\(ns)} & 
                            \makecell{Bytes\\Trans-\\fer} & 
                            \makecell{Packet\\Transfer\\(MTU\\1500)} & 
                            \makecell{Estimated\\PQC\\Strength} & 
                            \makecell{Estimated\\Classical\\Strength} \\
                            \hline
                            X25519-Kyber768-Draft00 & 1310505.00 & 1439010.02 & 13818200 & 926100 & 2336 & 2 & 192 & 128 \\
                            \hline
                            X25519-MLKE512M-Draft00 & 774315.00 &  138696.29 & 1648100 & 685400 & 1632 & 2 & 128 & 128 \\
                            \hline
                            X25519-MLKE768M-Draft00 & 951946.00 & 36843.45 & 1144500 & 907000 & 2336 & 2 & 192 & 128 \\
                            \hline
                            X25519-MLKE1024M-Draft00 & 1297202.00 & 77298.42 & 1831800 & 1232300 & 3200 & 4 & 256 & 128 \\
                            \hline
                            X448-MLKEM768-Draft00 & 3269694.00 & 492102.83 & 5326600 & 2990700 & 2384 & 2 & 192 & 224 \\
                            \hline
                            X25519e-FrodoKEM976-SHAKEDraft00 & 45021723.00 &  15273261.36 & \scriptsize 140257600 & 37245200 & 31440 & 22 & 192 & 128 \\
                            \hline
                            X25519-FrodoKEM976-SHAKEDraft00 & 44002536.00 & 18161708.88 & \scriptsize 179691800 & 37107400 & 31488 & 22 & 192 & 128 \\
                            \hline
                            X25519e-FrodoKEM976-AESDraft00 & 24082002.00 & 4538374.78 & 46741900 & 20640800 & 31440 & 22 & 192 & 128 \\
                            \hline
                            X25519-FrodoKEM976-AESDraft00 & 23845921.00 &  10479664.75 & 115511600 & 20615500 & 31488 & 22 & 192 & 128 \\
                            \hline
                            X25519-Mceliece-Draft00 & 84805784.00 & 25748678.22 & 214952000 & 59782600 & 200722 & 139 & 128 & 128 \\
                            \hline
                        \end{tabularx}
                    }
                    \label{tab:performance-framework}
                \end{table}
            \endgroup
            
    \section{Conclusion}
        This work presents a hybrid encryption framework that integrates EdDSA (Ed\-25519) and ECDH (X25519) for classical cryptography, ML-DSA-6x5 and ML-KEM-768 for post-quantum cryptography, and QKD keys managed by Guardian, delivering a quantum-resistant communication protocol. By combining these methods through a key derivation function, our prototype ensures confidentiality, integrity, and authenticity while leveraging Guardian’s scalable QKD key distribution across multi-node networks. Performance evaluations (Section \ref{performance-eval}) demonstrate the framework’s practicality, balancing classical efficiency with PQC’s quantum resilience. This approach not only bridges the transition from classical to quantum-resistant systems but also provides a deployable solution for organizations facing evolving quantum threats. Future work will try to explore real-world QKD hardware integration and optimize PQC performance for resource-constrained environments.

    \begin{credits}
        \subsubsection{\ackname} 
            We thank Goh Geok Ling, Goh Yue Jun, Lee Jun Quan, Tan Jia Ye, and Michael Kasper for their contributions, guidance, and support in this work.
            
        \subsubsection{\discintname}
            The authors declare that they have no conflict of interest.
    \end{credits}

    \bibliographystyle{splncs_citation_sort}
    \bibliography{references}

\end{document}